\documentclass[5p]{elsarticle}

\usepackage[english]{babel}
\usepackage{braket}
\usepackage{caption}
\usepackage{float}
\usepackage{graphicx}
\usepackage{bigints}
\usepackage{color}
\usepackage{amssymb}

\begin{document}

\begin{frontmatter}
\title{Singularity-free and non-chaotic inhomogeneous Mixmaster in polymer representation for the volume of the universe}
\author[sapienza,umd]{Stefano Antonini\corref{cor}}
\ead{stefanoantonini94@gmail.com, +1 301 683 4285}
\author[sapienza,enea]{Giovanni Montani}
\ead{giovanni.montani@enea.it}
\cortext[cor]{Corresponding author}
\address[sapienza]{Department of Physics, ``Sapienza'' University of Rome,
P.le Aldo Moro 5, 00185 Roma (Italy)}
\address[umd]{Department of Physics,
University of Maryland, College Park, MD 20742, USA}
\address[enea]{ENEA, Fusion and Nuclear Safety Department, C. R. Frascati,
Via E. Fermi 45, 00044 Frascati (Roma), Italy}

\date{\today}

\begin{abstract}
We analyze the semiclassical polymer dynamics of the inhomogeneous Mixmaster model by choosing the cubed scale factor as the discretized configurational variable, while the anisotropies remain pure Einsteinian variables. Such a modified theory of gravity should be regarded as the appropriate framework to describe the behavior of quantum mean values, both in polymer quantum mechanics and in Loop Quantum Cosmology. We first construct the generalized Kasner solution, including a massless scalar field and a cosmological constant in the dynamics. They account for a quasi-isotropization, inflationary-like mechanism. The resulting scenario links a singularity-free Kasner-like regime with a homogeneous and isotropic de Sitter phase. Subsequently, we investigate the role of the three-dimensional scalar curvature, demonstrating that a bounce of the point-universe against the potential walls can always occur within the polymer framework, also in the presence of a scalar field. However, the absence of the singularity implies that the curvature is bounded. Therefore, the point-universe undergoes an oscillatory regime until it oversteps the potential walls (if the Big Bounce is not reached before). After that, a final stable Kasner-like regime will last until the Big Bounce. Thus, the present study demonstrates that, as soon as a discretization of the volume of the universe is performed, the generic cosmological solution is non-chaotic and free from singularities. It is likely that the same result can be achieved also in the loop quantum cosmology approach.
\end{abstract}


\end{frontmatter}

\section{Introduction}

The construction of a ``generic'' inhomogeneous solution has represented one of the most important achievements of relativistic cosmology. Such a solution is valid asymptotically toward the initial singularity \cite{bkl82,kirillov93,montani95,beninimontani}, see also \cite{barrow,uggla}, and is obtained by extending the Mixmaster dynamics \cite{bkl70,misnerqc}, see also \cite{kirchaos,imponente2001,cornish}. Each point of space (de facto, each sufficiently small region around a space point, having the size of the average horizon) is described independently according to the so-called BKL conjecture \cite{review,primord}.

The analysis of the inhomogeneous oscillatory regime (``inhomogeneous Mixmaster'') emphasizes how the cosmological singularity is a very general feature of the Einstein equations. However, the non-physical and non-predictive nature of the Einstein equations when approaching the cosmological singularity suggests the necessity of a regularization of the theory. The most reliable picture of a non-singular universe has recently emerged from the implementation of loop quantum gravity \cite{canonic,rovelli} to the quantum dynamics of the isotropic universe \cite{ashtekar1,ashtekar2,ashtekar3}, see also \cite{cianfloop1,cianfloop2,marchini}.

Loop quantum gravity has been applied to various cosmological models \cite{loopbianchi1,loopbianchi2,loopbianchi3}. Possible implications on the BKL conjecture have been discussed in \cite{ashtekarbkl}. However, the loop quantization of the Mixmaster model presents non-trivial subtleties and cannot be regarded as fully achieved. Nonetheless, interesting results about the qualitative behavior of the model have been presented in \cite{wilson1,wilson2,bojo1,bojo2}, see also \cite{numerico1,numerico2,singh} for a numerical analysis, where it is argued that the homogenous Mixmaster model is, within the framework of loop quantum cosmology, a singularity-free universe. At least when a single contracting-expanding phase, i.e. a single Big Bounce, is considered, chaos is removed as well.

The non-trivial technical equipment required in loop quantization procedures has led to infer \cite{ashtekarpolymer1,ashtekarpolymer2,crino} that the so-called polymer quantum mechanics \cite{corichi1,corichi2} (i.e. quantum mechanics on a discrete lattice), when applied to the Minisuperspace, may provide results qualitatively equivalent to those of loop quantum cosmology. In this paper, we analyze the inhomogeneous Mixmaster model by adopting a semiclassical polymer approach, i.e. by replacing the standard Hamiltonian dynamics with the polymer one, within the prescriptions of the WKB limit. In this respect we remark how the polymer parameter $\mu$, which determines the discretization scale of the configurational variable, can be thought as independent of $\hbar$. Therefore, the semiclassical limit $\hbar\to 0$ with a fixed $\mu$ leads to a modified classical Einsteinian dynamics. This semiclassical analysis, completely analogous to that carried out for example in \cite{crino,corichi1,corichi2,lecian,mantero}, must be regarded as a qualitative and preliminary description of the behavior of the quantum mean values in the sense of the Ehrenfest theorem. Even if a natural further step will be the study of the full quantum theory, the reliability of the semiclassical polymer approximation in this context and its similarities with the semiclassical limit of loop quantum cosmology have been pointed out in the literature (see for instance \cite{loopbianchi1,ashtekarpolymer1,mantero,montoya}). Additionally, it is well known \cite{kirillov1997} that the Mixmaster model is essentially valid only in a classical or semiclassical regime, since in the full quantum scenario the piecewise character of the BKL conjecture is naturally lost due to the spreading of the universe wave function (see \cite{furusawa} for a numerical analysis of the Bianchi IX exact quantum dynamics). This fact, while justifying the use of a polymer semiclassical approach, allows also to retain the volume of the universe small enough for the BKL conjecture to hold before the
cutoff scale is approached.

Polymer quantization of the homogeneous Mixmaster model in standard Misner variables has been addressed both at a semiclassical and at a quantum level. By choosing the anisotropies as polymer configurational variables \cite{lecian}, a singular but non-chaotic cosmology is obtained. On the other hand, it is somewhat suprising that, when the polymer variable is the isotropic one \cite{crino}, both the singularity and the chaotic behavior of the Mixmaster are preserved. These two cases have been treated separately because they represent two physically different approaches. Indeed, by choosing the polymer representation for the anisotropies, we are affecting the real gravitational degrees of freedom, while the discretization of the isotropic variable, which is related to the volume of the universe, alter the geometrical properties of the system. Furthermore, the analysis carried out in \cite{mantero} shows how the dynamics of the isotropic universe changes under different choices of the polymer isotropic configurational variable. In particular, it has been demonstrated that the semiclassical results of loop quantum cosmology are achieved only by using the cubed cosmic scale factor, i.e. the volume of the universe, as the polymer variable. In this paper, we use an analogous isotropic variable to provide a semiclassical polymer description of the inhomogeneous Mixmaster.

After a brief review of the inhomogeneous Mixmaster and of polymer quantization, we will derive in Section \ref{nochaos} the Bianchi I polymer solution, including a scalar field and a cosmological constant in the dynamics. They allow to account for an inflationary scenario, able to isotropize the Mixmaster \cite{kirmont}. The absence of the cosmological singularity, replaced by a Big Bounce, will emerge naturally. Then, we will analyze the transition process between two successive Kasner regimes in the Misner representation \cite{misnerqc}. This study demonstrates that the bounces responsible for such transitions are always possible, surprisingly also in the presence of a scalar field. Afterward, we will remark how the absence of the singularity implies that the potential walls cannot be approximated as infinite anymore, but they have a well-defined maximum height. Hence, if the kinetic term of the ADM-reduced Hamiltonian is larger than the spatial curvature term, the point-universe is able to overstep the potential walls. We can infer that, after
a certain number of bounces, such a condition will be fulfilled, unless the Big Bounce is reached before. Indeed, the point-universe evolution explores the available phase space, according to the Mixmaster-like scenario, until it will be able to overstep the potential walls.
From that moment on, a final, stable Kasner-like regime will last until the Big Bounce. The chaotic behavior is suppressed. Clearly, if the Big Bounce is reached before the particle oversteps the potential walls, the chaos will be removed as well, due to the the finite number of bounces. This result is in accordance with the loop quantum cosmology analysis of the Mixmaster model \cite{wilson1,wilson2,bojo1,bojo2}, but generalizes the latter to a generic inhomogeneous universe. In fact, in Section \ref{gradients} we will prove that the BKL conjecture is still valid within the polymer framework. We remark that in our analysis the chaotic properties can be removed due to the polymer effects, that prevent the potential walls from becoming infinite and allow the point universe to overstep them, and not only because a finite number of bounces against the walls can occur before the Big Bounce. The chaos-removal mechanism based on the finite height of the potential wall is in accordance with the one in \cite{bojo1,bojo2}, where the effects of the inverse triads are taken into account, while the second mechanism, due to the presence of the Big Bounce and hence to the finite number of Kasner epochs, is analogous to the one discussed in \cite{wilson1,wilson2}. An interesting open question, as it is in \cite{wilson1,wilson2}, is whether the chaotic behavior is recovered when considering an infinite number of successive expanding and contracting phases, i.e. an infinite number of Big Bounces.

The present study possesses two main merits. First, it suggests the existence of a singularity-free and non-chaotic dynamics for a generic cosmological solution when the polymer quantization scheme is applied. A quasi-isotropization mechanism driven by the cosmological constant allows to link such an early phase of the universe with a later homogeneous and isotropic phase. Second, it emphasizes that these features, very similar to those present in loop quantum cosmology, are obtained only when the polymer quantization affects the dynamics of an isotropic variable directly related to a geometrical structure, i.e. the volume of the universe.

\section{Inhomogeneous Mixmaster}
\label{inho}

This section is devoted to characterizing the generic cosmological problem and its relationship with the homogeneous Mixmaster model. We will work in Planck units $c=\hbar=G=1$. The line element of a generic foliated spacetime is:
\begin{equation}
\textrm{d}s^2=N^2\textrm{d}t^2-h_{\alpha\beta}\left(\textrm{d}x^\alpha+N^\alpha
\textrm{d}t\right)\left(\textrm{d}x^\beta+N^\beta\textrm{d}t\right)
\label{line}
\end{equation} 
where $\alpha,\beta=1,2,3$, $N(t,\mathbf{x})$ is the lapse function and $N^\alpha(t,\mathbf{x})$ is the shift vector. Without loss of generality, the triadic projection of the three-dimensional metric tensor can be chosen to be diagonal:
\begin{equation}
\notag
h_{\alpha\beta}(t,\mathbf{x})=\textrm{e}^{q_1(t,\mathbf{x})}l^1_\alpha l^1_\beta+\textrm{e}^{q_2(t,\mathbf{x})}l^2_\alpha l^2_\beta+\textrm{e}^{q_3(t,\mathbf{x})}l^3_\alpha l^3_\beta.
\end{equation}
Following \cite{inoimp}, we assume the Kasner vectors $l^a_\alpha$ to depend only on spatial coordinates. It is important to remark that this picture is not the most general. Indeed, by this choice, we do not take into account the rotation of Kasner vectors, which has been proved to be a higher order dynamical effect \cite{primord}. The general case where Kasner vectors are allowed to rotate is described in \cite{beninimontani}. The generalized Misner variables are defined as:
\begin{equation}
\notag
q_a(t,\mathbf{x})=\frac{2}{3}\ln\left[V(t,\mathbf{x}) \right]+2\beta_a(t,\mathbf{x})
\end{equation}
with $a=1,2,3$ and $\beta_a=(\beta_++\sqrt{3}\beta_-,\beta_+-\sqrt{3}\beta_-,-2\beta_+)$. $V$ is the isotropic variable, proportional to the spatial volume of the universe, while $\beta_\pm$ are the anisotropies. Let us include a massless, self-interacting scalar field $\phi$ in the dynamics. Through a suitable symmetry-breaking mechanism, typical of the inflationary paradigm, the \textit{slow-roll} condition $\dot{\phi}^2\ll V(\phi)$ can be satisfied, as well as the condition:
\begin{equation}
\notag
\left|\nabla\phi\right|^2\ll V\left[\phi(t,\mathbf{x})\right]\sim \tilde{\Lambda}(\mathbf{x}).
\end{equation}
Rescaling the scalar field and by means of the Legendre transformation of the Einstein-Hilbert Lagrangian density associated to the line element (\ref{line}), the following Hamiltonian action functional is obtained:
\begin{equation}
\notag
\mathcal{S}=\int_{\mathbb{R}\otimes \Sigma_t}\textrm{d}^3x\textrm{d}t\left(p_\textrm{v}\frac{\partial V}{\partial t}+\sum_rp_r\frac{\partial \beta_r}{\partial t}-N\mathcal{H}-N^\gamma \mathcal{H}_\gamma\right)
\label{inaction}
\end{equation} 
where $r=+,-,\phi$ and $\beta_\phi\equiv\phi$. A variation with respect to the lapse function and the shift vector leads to the superhamiltonian and supermomentum constraints respectively:
\begin{equation}
\mathcal{H}=\frac{3\chi}{4}\left[-Vp_\textrm{v}^2+\frac{\sum_rp_r^2}{9V^2}+
\frac{V^{\frac{1}{3}}}{3\chi^2}U_{\textrm{in}}+V\Lambda(\mathbf{x})\right]=0
\label{superham}
\end{equation}
\begin{equation}
\begin{split}
\mathcal{H}_\gamma=&p_\textrm{v}\partial_\gamma V+\sum_rp_r\partial_\gamma\beta_r-\partial_\gamma\left(Vp_\textrm{v}+
\frac{1}{6}p_++\frac{\sqrt{3}}{6}p_-\right)\\
&+\frac{1}{6}\partial_\beta\left[2\sqrt{3}p_-l^\beta_2l^2_\gamma+(3p_+
+\sqrt{3}p_-)l^\beta_3l^3_\gamma\right]=0
\end{split}
\label{supermom}
\end{equation}
where $\chi=8\pi$ is the Einstein constant (in Planck units).
We remark how the self-interacting scalar field, under the slow-roll condition, is equivalent to a free scalar field and a cosmological constant $\Lambda(\mathbf{x})$. The dependence of $\Lambda$ on the spatial coordinates means that the slow-roll condition may be satisfied differently in different points of space. The potential term, due to the three-dimensional scalar curvature, can be split as $U_{\textrm{in}}=U_\textrm{B}+W$, where $W$ contains spatial gradients of the configurational variables and
\begin{equation}
U_B(\beta_+,\beta_-,\mathbf{x})=\sum_{a=1}^3\lambda_a^2\textrm{e}^{4\beta_a}
-\sum_{a\neq b}\lambda_a\lambda_b\textrm{e}^{\beta_a+\beta_b}
\label{potenziale}
\end{equation}
is the inhomogeneous generalization of the potential of the Bianchi models. The last term in equation (\ref{potenziale}) can be neglected in the limit $V\to 0$ \cite{inoimp}. Unlike the homogeneous case, the quantities $\lambda_a$ are not constant, but they are defined as $\lambda_a(\mathbf{x})=\textbf{l}_a(\mathbf{x})\cdot[\nabla\wedge \mathbf{l}_a(\mathbf{x})]/v(\mathbf{x})$ with $v(\mathbf{x})=\mathbf{l}_1(\mathbf{x})\cdot[\mathbf{l}_2(\mathbf{x})\wedge \mathbf{l}_3(\mathbf{x})]$. If the term $W$ in the potential is negligible with respect to $U_\textrm{B}$, the superhamiltonian reduces to that one of the Bianchi models, and the spatial coordinates appear only as parameters. Thus, by means of the gauge choice $N^\gamma=0$, each point of space evolves independently and can be described using a homogeneous Bianchi model \cite{bkl82}. The most general choice is to consider $\lambda_a(\mathbf{x})\neq 0$ $\forall a$, i.e. to examine models similar to either the Bianchi VIII or the Bianchi IX model, but where $\lambda_a$ are not necessarily unitary. Additionally, the corresponding solutions of the dynamics must satisfy also the supermomentum constraint, which identically vanishes for a homogeneous spacetime. By requiring $l_{in}\gg d_H\sim t$, where $l_{in}$ is the physical scale of the inhomogeneities and $d_H$ the average Hubble horizon, each causal connected region, instead of each point of space, can be regarded as a nearly homogeneous region, described by a Mixmaster model \cite{cianf}. In the following we will assume $W$ to be negligible according to the BKL conjecture \cite{bkl82}, approximating point-by-point the inhomogeneous cosmological model with a Mixmaster-like model. In Section \ref{gradients} we will verify the reliability of this assumption by solving the supermomentum constraint and explicitely evaluating the term $W$.

\section{Polymer quantization}

Polymer representation is a formulation of quantum mechanics unitarily-inequivalent to the Schr\"odinger one \cite{corichi1}. In order to build it, the starting point is the Weyl quantization procedure. A non-countable basis of the non-separable polymer kinematic Hilbert space $\mathcal{H}_{poly}$ is provided by abstract kets $\ket{\nu}$, with $\nu\in\mathbb{R}$. The scalar product between two fundamental kets is given by $\braket{\lambda|\nu}=\delta_{\lambda,\nu}$. A representation of the Weyl algebra on such a Hilbert space is obtained by means of two fundamental operators, the label operator $\hat{\epsilon}$ and the displacement operator $\hat{S}(\beta)$, whose action on the kets is:
\begin{equation}
\notag
\hat{\epsilon}\ket{\nu}=\nu\ket{\nu};\hspace{1cm}\hat{S}(\beta)\ket{\nu}=
\ket{\nu+\beta}.
\end{equation}
Since $\forall\beta\neq 0$ $\braket{\nu|\nu+\beta}=0$, the action of the displacement operator is discontinuous with respect to the parameter $\beta$. Thus, $\hat{S}(\beta)$ cannot be expressed in terms of an exponential of a Hermitian operator which generates the displacement. By considering an one-dimensional system with configurational variable $q$ and conjugate momentum $p$, the states corresponding to the fundamental kets in the momentum polarization take the form $\braket{p|\nu}=\textrm{e}^{\textrm{i}\nu p}$. Therefore, the operator $\hat{q}$ is identified with the label operator $\hat{\epsilon}$ and has a discrete spectrum (its eigenvalues are associated to orthogonal eigenstates). On the other hand, the displacement operator reads $\hat{S}(\beta)=\textrm{e}^{\textrm{i}\beta p}$ and the operator $\hat{p}$ is not defined. As a consequence, a generic Hamiltonian $H=p^2/(2m)+V(q)$ cannot be immediately promoted to be an operator on the Hilbert space. The definition of the dynamics requires the introduction of a lattice, called \textit{graph}, on the configurational space:
\begin{equation}
\notag
\gamma_\mu=\left\{q\in\mathbb{R}\big|q=n\mu, \forall n\in \mathbb{Z}\right\}
\end{equation}
where the polymer parameter $\mu$ is the distance between two neighboring points of the graph. In order to remain within this reduced configurational space, the parameter of the displacement operator must be a multiple of the polymer parameter: $\beta=k\mu$, $k\in\mathbb{Z}$. The fundamental displacement operator is then $\hat{S}(\mu)=\exp(\textrm{i}\mu p)$. By approximating the conjugate momentum as:
\begin{equation}
p\to\frac{1}{2\textrm{i}\mu}\left(\textrm{e}^{\textrm{i}\mu p}-\textrm{e}^{-\textrm{i}\mu p}\right)=\frac{\sin(\mu p)}{\mu}
\label{approx}
\end{equation}
it is now possible to promote the Hamiltonian to be an operator on the Hilbert space. The kinetic term will be in fact expressed by means of the fundamental displacement operator. The approximation (\ref{approx}) is clearly valid only if $\mu p\ll 1$. The problem of recovering the standard formulation of quantum mechanics (continuum limit) through successive refinements of the graph and a renormalization procedure is extensively discussed in \cite{corichi1,corichi2}. It is worth noting that, if one hypothesizes, as loop quantum gravity suggests, a space discretization at small length scales (for instance at the Planck scale), polymer representation can be regarded as a natural framework to study cosmological models. Indeed, the discrete nature of the space is reflected by the discrete structure of the polymer configurational space. Thus, the dynamical differences between the standard and the polymer-modified systems where the approximation (\ref{approx}) is not valid anymore may be a hint of the presence of new physics.

\section{Big Bounce and absence of chaos}
\label{nochaos}

In the following we will analyze the behavior of the cosmological model introduced in Section \ref{inho} from a semiclassical perspective. The superhamiltonian and supermomentum constraints will be modified by means of the approximation (\ref{approx}) and the corresponding classical Hamilton's equations will be solved. The discrete configurational variable is the isotropic variable $V$. This study can be regarded as a zeroth order WKB approximation of the full quantum system. By neglecting spatial gradients and choosing $0<\lambda_a\leq 1$ $\forall a$ in the potential term (\ref{potenziale}), the polymer-modified superhamiltonian constraint reads:
\begin{equation}
\mathcal{H}=\frac{3\chi}{4}\left[-V\frac{\sin^2(\mu p_\textrm{v})}{\mu^2}+\frac{\sum_rp_r^2}{9V}+
\frac{V^{\frac{1}{3}}}{3\chi^2}U_{IX}+V\Lambda\right]=0
\label{superham9}
\end{equation}
where $U_{IX}$ is the Mixmaster-like potential. Let us first investigate the behavior of the model in a region where the potential term is negligible. The superhamiltonian constraint imposes
\begin{equation}
\sin^2(\mu p_\textrm{v})=\mu^2\left(\frac{\sum_rp_r^2}{9V^2}+\Lambda\right).
\label{vincolo}
\end{equation}
Thus, the superhamiltonian constraint can be satisfied only if the right hand side of equation (\ref{vincolo}) is smaller than 1. The latter condition leads to a lower bound for the volume of the universe:
\begin{equation}
V>V_{bounce}\equiv\frac{\mu}{3}\sqrt{\frac{\sum_rp_r^2}{1-\mu^2\Lambda}}.
\label{vmin}
\end{equation}
In other words, in the polymer framework the volume cannot vanish and the singularity is not present anymore. It is replaced by a Big Bounce. The condition $\mu^2\Lambda<1$, necessary for the minimum volume $V_{bounce}$ to be defined, is physically reasonable. Indeed, in Planck units $L_p=E_p=1$. We note how the minimum volume is determined not only by the polymer discretization scale $\mu$, but also by the conjugate momenta $p_r$ (which are constants of motion when the potential term is negligible) and by the cosmological constant. We remark that, by a suitable choice of the initial conditions, the Big Bounce takes place when the volume of the universe is much larger than the discretization scale. This feature is in accordance with the results obtained in loop quantum cosmology \cite{numerico1,numerico2}.

The Hamilton's equation for the volume $V$ reads:
\begin{equation}
\frac{dV}{dt}=-\frac{3\chi}{2}NV\frac{\sin(\mu p_\textrm{v})\cos(\mu p_\textrm{v})}{\mu}.
\label{hamv}
\end{equation}
It suggests that, in order to deal with an expanding universe, we must choose one of the following branches for the momentum $p_\textrm{v}$:
\begin{equation}
\notag
\frac{n\pi}{\mu}-\frac{\pi}{2\mu}<p_\textrm{v}<\frac{n\pi}{\mu},\hspace{0.5cm} n\in\mathbb{Z}.
\label{branches}
\end{equation}
The natural choice is $n=0$, which gives the correct continuum limit $\mu\to 0$. With this choice, in the synchronous gauge ($N=1$) and using the superhamiltonian constraint (\ref{vincolo}), equation (\ref{hamv}) can be rewritten as:
\begin{equation}
\begin{split}
\frac{dV}{dt}&=\frac{\chi}{2}\sqrt{\sum_rp_r^2+9\Lambda V^2}\sqrt{1-\frac{\mu^2}{9V^2}\left(\sum_rp_r^2+9\Lambda V^2\right)}\\
&=\frac{\chi}{2}\frac{A}{V}\sqrt{\left(\frac{BV^2+C}{A}\right)^2-1}
\end{split}
\label{hamv2}
\end{equation}
where we defined
\begin{equation}
\notag
\begin{aligned}
&A=\frac{\sum_rp_r^2}{3}\sqrt{\mu^2+\frac{1}{4\Lambda}\frac{(1-2\mu^2\Lambda)^2}{1-\mu^2\Lambda}}\\
&B=3\sqrt{\Lambda}\sqrt{1-\mu^2\Lambda}\\
&C=\frac{\sum_rp_r^2}{6\sqrt{\Lambda}}\frac{1-2\mu^2\Lambda}{\sqrt{1-\mu^2\Lambda}}.
\end{aligned}
\end{equation}
Using the change of variable
\begin{equation}
\notag
x=\frac{BV^2+C}{A}
\end{equation}
the solution of equation (\ref{hamv2}) reads:
\begin{equation}
\notag
\cosh^{-1}(x)=\chi Bt+C_1.
\end{equation}
It is immediate to check that $x|_{V=V_{min}}=1$. Therefore, by means of the initial condition $V(t=0)=V_{bounce}$, i.e. $x(t=0)=1$, we find $C_1=0$. The expression of the volume of the universe as a function of the synchronous time is finally:
\begin{equation}
V(t,\mathbf{x})=\sqrt{\frac{\sum_rp_r^2\left[\cosh\left(3\chi\sqrt{\Lambda}
\sqrt{1-\mu^2\Lambda}\hspace{0.1cm} t\right)-1+2\mu^2\Lambda\right]}{18\Lambda(1-\mu^2\Lambda)}}.
\label{volume}
\end{equation}
\begin{figure}[H]
\centering
\includegraphics[width=\columnwidth]{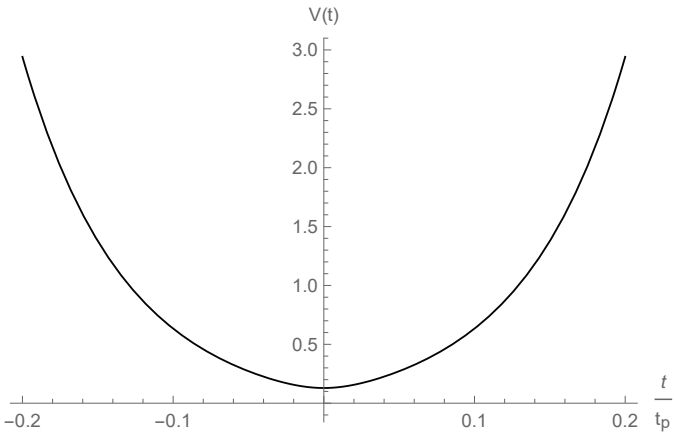} 
\caption{Volume $V(t,\mathbf{x})$ (extended to negative values of $t$) as a function of the synchronous time $t$. $p_r=\Lambda=0.2$, $\mu=1$. A choice for $\Lambda$ consistent with inflationary theories would be several order of magnitude smaller. $\Lambda=0.2$ allows to emphasize the behavior of the bridge solution.}
\label{figvol}
\end{figure}
The singularity is clearly replaced by a Big Bounce. The Bounce occurs when the volume reaches the minimum volume given by equation (\ref{vmin}). Far from the Bounce, the volume grows exponentially due to the presence of the cosmological constant. By performing the ADM reduction of the system and choosing the volume $V$ as the internal time variable ($\dot{V}=1$), the solution of the Hamilton's equations associated to the anisotropies and to the scalar field can be written as:
\begin{equation}
\beta_r(V,\mathbf{x})=\frac{p_r}{3\sqrt{\sum_rp_r^2}}\left[\mathcal{G}\left(
V,\mu,\Lambda,p_r\right)+\ln\left(\frac{\mu}{6}\right)\right]+C_{\beta_r}(\mu)
\label{ani}
\end{equation}
with
\begin{equation}
\mathcal{G}\equiv\frac{1}{\sqrt{\Lambda}\mu}\Re\left\{F\left[\textrm{i}
\sinh^{-1}\left(3\sqrt{\frac{\Lambda}{\sum_rp_r^2}}V\right)
\Bigg|1-\frac{1}{\Lambda\mu^2}\right]\right\}
\label{ellipt}
\end{equation}
where $\Re$ indicates the real part and $F(x|m)$ is the incomplete elliptic integral of the first kind. The logarithmic term is obtained by splitting the constant of integration and allows to recover the standard (non-polymer) solution of the Hamilton's equations in the limit $\mu\to 0$ \cite{kirmont}. By substituting expression (\ref{volume}) for the volume into the solution (\ref{ani}), the behavior of the anisotropies as functions of the synchronous time can be investigated.
\begin{figure}[H]
\centering
\includegraphics[width=\columnwidth]{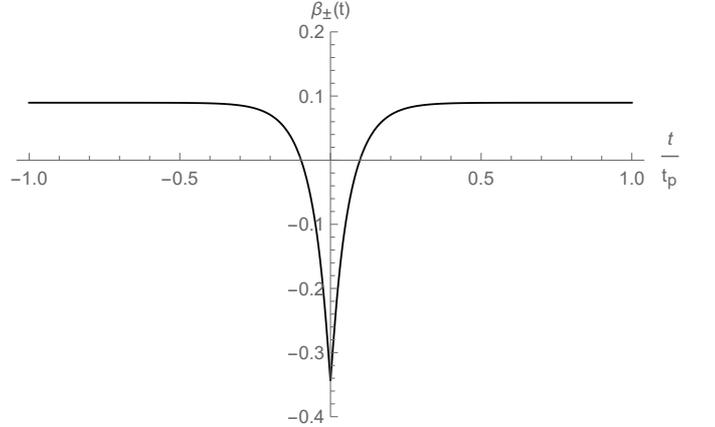} 
\caption{Anisotropies $\beta_{\pm}(t,\mathbf{x})$ (extended to negative values of $t$) as a function of the synchronous time $t$. $p_r=\Lambda=0.2$, $\mu=1$, $C_{\beta_r}=0$. By choosing $p_r<0$ the anisotropies reach their maximum value at the Big Bounce.}
\label{figani}
\end{figure}
In the polymer scenario, approaching the Big Bounce, the anisotropies do not diverge as it happened in the standard case. Indeed, when $V=V_{bounce}$ the term (\ref{ellipt}) vanishes and they assume a constant value logarithmically dependent on the polymer parameter $\mu$, which acts as a cutoff. On the other hand, far from the Bounce, the anisotropies tend toward a constant value. The latter can be absorbed by an appropriate (and local) rescaling of the coordinate system. We obtained a \textit{bridge solution}, connecting a non-singular, inhomogeneous and anisotropic, Kasner-like early phase of the universe with a later homogeneous and isotropic phase. Thus, the quasi-isotropization mechanism described in \cite{cianf,kirmont} is preserved under the polymer modification. It is remarkable that, if the Kasner vector $\mathbf{l}_a$ associated to the scale factor growing toward the singularity satisfies the condition $\mathbf{l}_a\cdot(\nabla\wedge \mathbf{l}_a)=0$, this solution is stable toward the Big Bounce \cite{bklin}. It is important to note here that this result is local, and since it is derived in the WKB scenario, isotropization is obtained only in a single WKB patch, i.e. in a nearly homogeneous region. Nonetheless, under the assumptions necessary for considering the inhomogeneous Mixmaster model, i.e. $l_{in}\ll d_H$, such a region is larger than the average cosmological horizon \cite{cianf,kirmont}. Therefore, each causal connected region undergoes such an isotropization phase and can be expanded by the inflationary mechanism on a scale much larger than the Hubble radius. Thus, if the inflationary phase is long enough, it is possible to obtain a homogeneous and isotropic region larger than the Hubble horizon today. 

As it is well-known \cite{misnerqc}, the ADM-reduced system is equivalent to the description of a two-dimensional particle (the so-called point-universe) with coordinates $\beta_+$ and $\beta_-$, while $V$ is regarded as the time variable. The polymer-modified ADM Hamiltonian reads:
\begin{equation}
H_{ADM}=\frac{1}{\mu}\arcsin\left[\frac{\mu}{3V}\sqrt{\sum_rp_r^2+
\frac{3}{\chi^2}V^{\frac{4}{3}}U_{IX}+9\Lambda V^2}\right].
\label{admham}
\end{equation}
Following the same procedure carried out in \cite{misnerqc}, in the standard (non-polymer) description the last term in equation (\ref{potenziale}) can be neglected and the potential term can be regarded point-by-point as a triangular infinite and vertical wall when approaching the singularity. When the ``particle'' representing the universe is far from the walls, it behaves according to the Bianchi I dynamics. The walls recede with a speed $\left|\beta'_{wall}\right|=1/(6V)$. In vacuum the speed of the particle is $\left|\beta'\right|=1/(3V)$. Therefore it always reaches the walls and bounces against them, no matter its angle of incidence. The bounce corresponds to the transition between two Kasner epochs, i.e. to the change of the momenta $p_\pm$. Since the speed of the particle diverges when approaching the singularity, an infinite number of bounces takes place and the system shows chaotic and ergodic properties \cite{kirchaos,lifshitzcaos}. $p_\pm$ will assume all the possible values. The presence of a scalar field suppresses this chaotic behavior, ensuring the existence of a final, stable Kasner epoch \cite{BKscalarfield}. In the polymer-modified scenario, the potential wall is not infinite anymore. Indeed, the divergence of the potential wall corresponds to the divergence of the three-dimensional scalar curvature, occuring in the limit $V\to 0$. Since in our model the volume cannot vanish, the potential wall is always bounded, and it will reach its maximum height at the Big Bounce, i.e. at $V=V_{bounce}$, as we will show. Nonetheless, as long as the kinetic term in the ADM Hamiltonian is smaller than the potential one, the walls can still be considered infinite. 

We can therefore use the Misner procedure \cite{misnerqc} to find the speed of the walls. In particular, taking into account only the left wall, its position can be found by requiring the potential term to be of the same order of the potential-free (Bianchi I) ADM Hamiltonian:
\begin{equation}
\frac{\mu^2\lambda_3^2}{3\chi^2}\frac{\textrm{e}^{-8\beta_{wall}}}{V^{\frac{2}{3}}\sin^2(\mu H_{ADM}^I)}\sim 1.
\label{velmuri}
\end{equation}
By neglecting constant terms, the condition (\ref{velmuri}) can be rewritten as
\begin{equation}
\beta_{wall}\sim \frac{1}{6}\ln (V)-\frac{1}{4}\ln\left[\frac{3V}{\mu}\sin\left(\mu H_{ADM}^I\right)\right].
\label{posmuro}
\end{equation}
The derivative of equation (\ref{posmuro}) with respect to $V$ gives the speed of the wall:
\begin{equation}
\left|\beta'_{wall}\right|=\frac{1}{6V}-\frac{9\Lambda V}{4\left(\sum_rp_r^2+9\Lambda V^2\right)}.
\label{velmuro}
\end{equation}
Close to the Big Bounce, if $V_{bounce}\ll 1$ (we will see that we need such a condition for the validity of the present argument), then $V\ll 1$ and the second term of the RHS of equation (\ref{velmuro}) is negligible with respect to the first one. Thus, the speed of the wall is again $\left|\beta'_{wall}\right|=1/(6V)$. The analysis for the other two potential walls leads to the same result.

The speed of the particle in the presence of a scalar field can be found in the same way as in the standard case as well. The Hamilton's equations for the anisotropies can be written as:
\begin{equation}
\notag
\frac{d\beta_\pm}{dV}=\frac{p_\pm}{9V^2}\frac{\mu}{\sin\left(\mu H_{ADM}\right)\cos\left(\mu H_{ADM}\right)}
\end{equation}
and the speed of the point universe reads:
\begin{equation}
\notag
\left|\beta'\right|=\frac{1}{3V\cos\left(\mu H_{ADM}\right)}\sqrt{\frac{\mu^2(p_+^2+p_-^2)}{9V^2\sin^2\left(\mu H_{ADM}\right)}}.
\end{equation}
Using the expression for $p_+^2+p_-^2$ obtained from equation (\ref{admham}) and remembering that $\left.\sin\left(\mu H_{ADM}\right)\right|_{V=V_{bounce}}=1$, the speed of the particle when approaching the Big Bounce and far from the potential walls is finally given by:
\begin{equation}
\left|\beta'\right|=\frac{1}{3V\cos\left(\mu H_{ADM}\right)}\sqrt{1-\frac{p_\phi^2+9\Lambda V_{bounce}^2}{\sum_rp_r^2+9\Lambda V_{bounce}^2}}.
\label{velpart}
\end{equation}
We remark that, since $\left.\cos\left(\mu H_{ADM}\right)\right|_{V=V_{bounce}}=0$, the speed of the particle diverges when $V\to V_{bounce}$, while $\left|\beta'_{wall}\right|$ is finite. Hence, unlike the standard case, the particle reaches the wall for any angle of incidence even in the presence of a scalar field. Therefore, bounces will occur and, in analogy with the standard case, $p_\pm$ will assume all the possible values, until the condition
\begin{equation}
\sum_rp_r^2+9\Lambda V^2>\max\left(\frac{3}{\chi^2}V^{\frac{4}{3}}U_{IX}\right)
\label{cond1}
\end{equation}
will be satisfied, unless the Big Bounce is reached before. Indeed, since we have shown that the point-universe can always reach the potential wall (before the bounce), in analogy with the standard Einsteinian dynamics \cite{misnerqc}, we can infer the existence of a stochastic behavior of the system and, in particular, of the discrete map in the generalized Bianchi I parameter space (see for instance the analysis of the generalized BKL map developed in \cite{crino}, where polymer features are addressed). For a discussion of the chaotic properties of the  generic inhomogeneous cosmologies in the framework of continuous dynamical fluxes, see \cite{beninimontani,imponente2001}. Thus, we can argue that all the possible values for the parameters (Bianchi I constants of motion) $p_{\pm}$ could in principle be explored during the iteration of the point-universe bounces. As a result, when these quantities take values corresponding to the validity of relation (\ref{cond1}), we obtain the emergence of a free motion until the Big Bounce, and the chaotic features of the solution are suppressed. This is exactly the same argument adopted in \cite{BKscalarfield} to prove the suppression of the oscillatory regime due to the presence of a massless scalar field. From a dynamical system point of view, the point-universe invades all the allowed phase space and de facto chaos disappears because the configurational domain is no longer dynamically closed. Clearly, it can happen that the Big Bounce occurs before condition (\ref{cond1}) is satisfied. Thus, in both cases chaos is removed. When condition (\ref{cond1}) is fulfilled before the Big Bounce, the mechanism of chaos suppression is analogous to the one discussed in \cite{bojo1,bojo2}. Alternatively, if the Big Bounce is reached before (\ref{cond1}) holds, the chaos disappears due to the finite number of bounces of the point universe against the potential walls, in accordance with the analysis of \cite{wilson1,wilson2}.

Let us take into account only the left wall of the potential (associated to the term $\lambda_3^2\textrm{e}^{-8\beta_+}$). In order to estimate the height of the potential wall, we can follow the same reasoning described by Misner in \cite{misnerqc}, and use the modified Kasner-type expression (\ref{ani}) for the anisotropies. Indeed, in the classical picture, even if such a solution is valid only in absence of potential, the verticality of the potential wall guarantees that it is valid arbitrarily close to the wall. In the polymer scenario, the wall is not exactly vertical, but this approximation is still reliable as long as the slope of the potential wall is steep enough, i.e. as long as we have a small minimum volume $V_{bounce}\ll 1$. By substituting (\ref{ani}), the right hand side of the condition (\ref{cond1}) has a maximum in $V=V_{bounce}$. The condition can then be rewritten as:
\begin{equation}
p_+^2+p_-^2>\frac{3\lambda_3^6(1-\mu^2\Lambda)\textrm{e}^{-24C_{\beta_+}}}
{6^{-8\cos\theta_i}\chi^6}\mu^{4(1-2\cos\theta_i)}-p_\phi^2
\label{cond2}
\end{equation}
where $\cos\theta_i=p_+/\sqrt{p_+^2+p_-^2}$ is the angle of incidence of the particle. It is worth noting that if the particle approaches the left wall, then $\theta_i<\pi/3$. If $\theta_i>\pi/3$, it is possible to find an analogous condition for another wall. The compatibility of condition (\ref{cond2}) with the condition $V_{bounce}\ll 1$ is checked in Section (\ref{gradients}). When the polymer parameter $\mu$ is finite, the right hand side of expression (\ref{cond2}) is always bounded and, after a number of bounces, the condition can be satisfied. When it happens, the particle will overstep the potential wall and a final, stable Kasner epoch will last until the Big Bounce. The minimum volume will be that associated to the last Kasner epoch. The same reasoning is valid also in vacuum. In the continuum limit $\mu\to 0$, the right hand side of the condition (\ref{cond2}) diverges and we recover the infinite walls of the standard description. We finally remark that, since the Mixmaster expands, reaches a turning point and then recollapses \cite{linwald}, if the accelerated expansion comes to an end, the presence of the Big Bounce suggests that our model describes a cyclical universe.

We will now briefly focus our attention on the choice of the volume variable $V$ as clock for the dynamics. First of all, since our analysis is based on Hamilton's	equations, i.e. it addresses a modified but classical gravitational dynamics in the spirit of the Ehrenfest theorem, the specification of a time variable is essentially equivalent to any other choice, since such a variable is always classically linked to the synchronous time. The situation is significatively different when the full quantum analysis is carried out, because the classical link to the coordinate time is lost. As an example of such a difference, we remark how using the scalar field as a quantum clock is a very peculiar choice \cite{mantero}, see also \cite{bomba}. Indeed, it leads to a divergence of the cosmic scale factor of the isotropic universe for a finite value of the scalar field. The singular behavior is clearly removed when passing, on a classical footing, to the synchronous time, but quantumly it is not avoidable.

These peculiarities are, in general, due to the behavior of the derivative of the adopted time with respect to the synchronous one or to another internal time. In particular, we note that the presence of the Big Bounce implies that the variable $V$ is no longer monotonic as a function of the synchronous time, and its derivative $dV/dt$ vanishes when approaching the Big Bounce. The most remarkable consequence of such a feature is the divergence of the speed of the point-universe (\ref{velpart}). Using the Hamilton's equation for the scalar field, it is immediate to verify that, by choosing, for instance, the scalar field as the internal time variable, the speed of the point-universe at the Big Bounce is given by $\left|\beta'\right|=\sqrt{p_+^2+p_-^2}/p_\phi$ (which is a constant of motion), while the speed of the wall vanishes. Thus the chaos-removal mechanism previously analyzed is preserved. However, it is worth to emphasize that, using $\phi$ as the time variable, it clearly turns out that the number of bounces of the point universe against the potential walls is finite (because the range of $\phi$ is also finite, since it behaves exactly as the anisotropies $\beta_\pm$ (\ref{ani})). At a semiclassical level, this feature is clearly valid also using $V$ as the time variable, despite the divergence of the speed of the particle. Indeed, it can be verified from equation (\ref{velpart}) that $\left|\beta'\right|\propto 1/\sqrt{V-V_{bounce}}$, while the time interval before the Big Bounce is $\Delta=V-V_{bounce}$.

Nonetheless, in the quantum case it is fundamental and non-trivial to choose a specific internal time. The use of the time variable $V$ is the most natural when studying the evolution of the point-universe during the Mixmaster dynamics, because it is directly involved in the behavior of the potential term. Clearly, the result concerning the chaotic or non-chaotic properties of the inhomogeneous Mixmaster cannot depend, at least at this semiclassical level, on  the choice of a specific clock. The problem would be much more serious in the full quantum picture, since the equivalence of the different choices of the clock could be no longer valid.

Finally, we stress that, on a quantum level, the request of a monotonic behavior of the considered chosen time with respect to the label time of the spacetime foliation is a good requirement for a relational quantum time \cite{isham,timeqg}. Additionally, the physical intuition of the concept of time suggests that the expanding and the collapsing branches of the evolution of the universe should be regarded as separated phases on a quantum level. This point of view is consistent with the idea that expanding and contracting solution modes could correspond to a frequency separation (if possible) in the canonical quantum procedure. This is immediate to recognize for the Bianchi I model using the standard Misner variable, since its Wheeler-DeWitt equation is isomorphic to a massless Klein-Gordon equation.

\section{Spatial gradients}
\label{gradients}

\begin{figure}[H]
\centering
\includegraphics[width=\columnwidth]{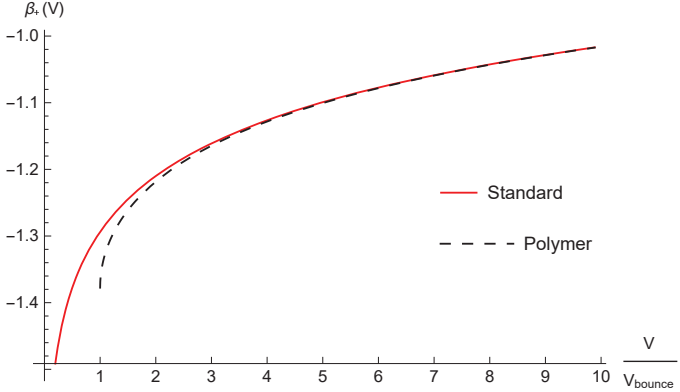} 
\caption{Standard and polymer anisotropy $\beta_+$ as a function of the volume $V$. $p_+=1.1$, $p_-=p_\phi=2$, $\lambda_3=1$, $\Lambda=0.2$, $\mu=0.0001$, $C_{\beta_r}(\mu)=-0.0482$, $C_{\beta_r}^{class}=0$. With this choice, the conditions (\ref{cond3}) are fulfilled and $V_{bounce}=0.000101\sim\mu$.}
\label{fig3}
\end{figure}

\begin{figure}[H]
\centering
\includegraphics[width=\columnwidth]{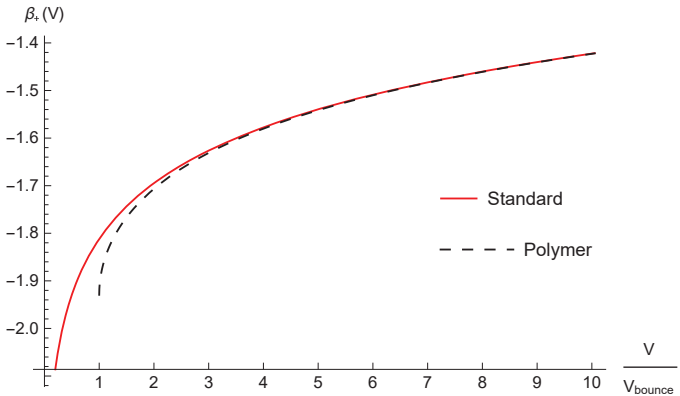} 
\caption{Standard and polymer anisotropy $\beta_+$ as a function of the volume $V$. $p_+=50$, $p_-=p_\phi=60$, $\lambda_3=1$, $\Lambda=0.2$, $\mu=0.0001$, $C_{\beta_r}(\mu)=-0.0676$, $C_{\beta_r}^{class}=0$. With this choice, the conditions (\ref{cond3}) are fulfilled and $V_{bounce}=0.0033\gg\mu$.}
\label{fig4}
\end{figure}

In the previous section we have assumed the BKL conjecture to be valid also in polymer representation. Thus we have neglected both the last term in equation (\ref{potenziale}) and the term $W$ in the potential, which contains spatial gradients of the configurational variables. Let us now justify this assumption. Starting from the components of the three-dimensional scalar curvature, obtained in \cite{bklin}, and using the generalized Misner variables, it is possible to write the explicit expression for $W$. All the terms contained in $W$ have a factor $\exp(\beta_a+\beta_b)$ with $a\neq b$ and $\beta_a=(\beta_++\sqrt{3}\beta_-,\beta_+-\sqrt{3}\beta_-,-2\beta_+)$, and some of them depend on spatial gradients of the configurational variables ($\partial_\gamma V$, $\partial_\gamma \beta_\pm$). The supermomentum constraint (\ref{supermom}) must be modified according to the polymer prescription (\ref{approx}). Then, relying on the same reasoning outlined in the previous section, if $V_{bounce}\ll 1$ we can use the expression (\ref{ani}) for the anisotropies and solve the supermomentum constraint to obtain an explicit expression for $\partial_\gamma V$, which does not diverge in the limit $V\to V_{bounce}$, and is convergent also in the continuum limit $\mu\to 0$. Additionally, evaluating $\partial_\gamma \beta_\pm$ by means of expression (\ref{ani}) for the anisotropies and substituting $\partial_\gamma V$, in the limit $V\to V_{bounce}$ we obtain:
\begin{equation}
\notag
\partial_\gamma \beta_\pm=\frac{\partial_\gamma p_\pm}{3\sqrt{\sum_rp_r^2}}\ln\left(\frac{\mu}{6}\right)-
\frac{p_\pm\sum_rp_r\partial_\gamma p_r}{3\left(\sum_rp_r^2\right)^\frac{3}{2}}\ln\left(\frac{\mu}{6}\right)+K
\end{equation}
where
\begin{equation}
\notag
\begin{split}
K\left(p_r,\partial_\gamma p_r, \textbf{l}_a\right)&=\frac{p_\pm}{6\sum_rp_r^2}\left\{\partial_\gamma p_+ + \sqrt{3}\partial_\gamma p_-\right.\\
&\left.-\partial_\beta\left[2\sqrt{3}p_-l^\beta_2l^2_\gamma+(3p_+
+\sqrt{3}p_-)l^\beta_3l^3_\gamma\right]\right\}.
\end{split}
\end{equation}
Therefore, $\partial_\gamma \beta_\pm$ is finite when $\mu\neq 0$. These spatial gradients depend on the polymer parameter $\mu$ logarithmically, thus they are logarithmically divergent in the continuum limit. Such a logarithmic behavior confirms, also in the polymer framework, the estimate provided in \cite{kirgrad}. It is well-known \cite{inoimp} that, in the standard picture and in the limit $V\to 0$, the terms $\exp(\beta_a+\beta_b)$ with $a\neq b$ are negligible with respect to $\exp(2\beta_a)$ where the potential is relevant. Thus, in the polymer framework, they are negligible if the condition $V\ll 1$ is satisfied. Therefore, the condition $V_{bounce}\ll 1$ must be satisfied too, because $V_{bounce}\leq V$. We observe how this condition is exactly the same we needed in order to use expression (\ref{ani}) for the anisotropies in the estimate of the spatial gradients here and of the height of the potential wall in Section \ref{nochaos}. It is remarkable that, if $V_{bounce}\ll 1$ is satisfied for the first Kasner epoch, it is satisfied for all the following Kasner epochs. Indeed, $V_{bounce}$ cannot become bigger than $V$, which decreases monotonically. Since the logarithmic growth of the spatial gradients cannot affect this behavior, if $V_{bounce}\ll 1$ both the last term in the Bianchi-like potential (\ref{potenziale}) and the term $W$ are negligible. Therefore, by requiring $V_{bounce}\ll 1$, i.e. $\sum_rp_r^2\ll 9(1-\mu^2\Lambda)/\mu^2$, the BKL picture holds and the analysis carried out in the previous section is completely justified. The last step is to verify that such a condition, which must be always satisfied, is compatible with the condition (\ref{cond2}). Both conditions are satisfied if the relation
\begin{equation}
\frac{3\lambda_3^6(1-\mu^2\Lambda)\textrm{e}^{-24C_{\beta_+}}}
{6^{-8\cos\theta_i}\chi^6}\mu^{4(1-2\cos\theta_i)}<\sum_rp_r^2\ll
\frac{9(1-\mu^2\Lambda)}{\mu^2}
\label{cond3}
\end{equation}
holds. It is possible only if (we recall that, since we are considering only the left wall, $\theta_i<\pi/3$)
\begin{equation}
\begin{cases}
\mu\ll \left(\frac{3\chi^66^{-8\cos\theta_i}}{\lambda_3^6\textrm{e}^{-24C_
{\beta_+}}}\right)^\frac{1}{2(3-4\cos\theta_i)}\hspace{1cm} \frac{1}{2}<\cos\theta_i<\frac{3}{4}\\
\mu\gg \left(\frac{3\chi^66^{-8\cos\theta_i}}{\lambda_3^6\textrm{e}^{-24C_
{\beta_+}}}\right)^\frac{1}{2(3-4\cos\theta_i)}\hspace{1cm} \frac{3}{4}<\cos\theta_i<1.
\end{cases}
\label{cond4}
\end{equation}
If the conditions (\ref{cond4}) are satisfied, it is possible to fulfill condition (\ref{cond3}) and the generic cosmological solution behaves as outlined in Section \ref{nochaos}. We remark how there exists a range of parameters, i.e. of initial conditions on the last Kasner epoch, such that also the condition $V_{bounce}\gg\mu$ is satisfied. This result recalls those obtained in loop quantum cosmology \cite{numerico1,numerico2}. As an example, we plot in Figures \ref{fig3} and \ref{fig4} the polymer-modified anisotropy $\beta_+$ (see equation (\ref{ani})) and its standard (classical) counterpart as a function of the volume for different choices of the parameters. In both cases, the conditions (\ref{cond3}) are satisfied. The expression for the classical anisotropies, solution of the classical Hamilton's equations, is given by:
\begin{equation}
\notag
\beta_r^{class}\left(V,\mathbf{x}\right)=\frac{p_r}{3\sqrt{\sum_rp_r^2}}\ln\left[\frac{\sqrt{\frac{9\Lambda}{\sum_rp_r^2}}V}{1+\sqrt{1+\frac{9\Lambda}{\sum_rp_r^2}V^2}}\right]+C_{\beta_r}^{class}.
\end{equation}
In Figure \ref{fig3} our choice of the initial conditions implies that $V_{bounce}\sim \mu$, while in Figure \ref{fig4} $V_{bounce}\gg \mu$. We remark the similarity of the behavior in the two cases: the polymer-modified anisotropy behaves differently from the standard one very close to the Big Bounce, reaching its minimum value for $V=V_{bounce}$, but when $V\sim 4 V_{bounce}$ the classical and the polymer solutions are already almost indistinguishable.
Since, as we have showed, spatial gradients only grow logarithmically in the polymer parameter, they cannot become dominant and destroy the inhomogeneous Mixmaster approximation. In particular, in our model sub-horizon spikes \cite{spike1,spike2,spike3,spike4} are not present. This may be a hint that the regularization due to the presence of a cutoff in the polymer quantization framework suppresses spikes. Alternatively, the absence of spikes could be related to our choice of non-rotating Kasner vectors. Indeed, even if the rotation of Kasner vectors is a higher order dynamical effect, we cannot exclude that it could be the source of sub-horizon sized spikes. Therefore, they may be present when the most general case \cite{beninimontani} is considered. A detailed analysis of this topic, although interesting, resides outside the goals of this paper.

\section{Concluding remarks}

In this work we have constructed the generic cosmological solution in the semiclassical polymer representation for the isotropic variable only. The presence of a Big Bounce naturally arises when choosing the cubed scale factor as the generalized coordinate, while the chaotic behavior of the (inhomogeneous) Mixmaster is suppressed. These results significantly overlap those obtained in loop quantum cosmology, developed in \cite{wilson1,wilson2,bojo1,bojo2,numerico1,numerico2}. Moreover, the flexibility of the polymer scenario has allowed us to extend the analysis to the generic inhomogeneous case. This suggests that also loop quantum cosmology would produce a generic non-chaotic bounce cosmology when applied to the cosmological problem with no symmetry restrictions \cite{cianfloop1,cianfloop2,ashtekarbkl}.

The present study (similarily to \cite{mantero}) also strongly suggests that the polymer quantization in the Minisuperspace is equivalent to loop quantum cosmology only if the discretized variable is directly related to a geometrical quantity. In fact, as shown in \cite{crino}, by discretizing the standard isotropic Misner variable, i.e. the logarithm of the volume instead of the volume itself, both the singularity and the chaotic behavior of the Mixmaster model are still present, and their properties are analogous to the standard, non-polymer case.

Apart from the cosmological relevance of the generic inhomogeneous solution we have constructed, which is able to link a non-singular Kasner-like dynamics with the isotropic de Sitter phase, the dependence of the behavior of the model on the choice of the discretized configurational variable and the analogies with the loop quantum cosmology approach should be regarded as the main results of the present work.

\bibliography{references}


\end{document}